\documentclass[aps,nofootinbib,showpacs,twocolumn]{revtex4}

\usepackage{graphicx}
\usepackage{amssymb}
\usepackage{amsmath}
\usepackage[pdfborder={0 0 0}]{hyperref}

\begin{document}

\title{Nonextensive local composition models in theories of solutions}

\author{Ernesto P. Borges}
\email{ernesto@ufba.br}

\affiliation{Instituto de Fisica, Universidade Federal da Bahia, BA 40210-340, Brazil \\
National Institute of Science and Technology for Complex Systems, Brazil}

\begin{abstract}
Thermodynamic models present binary interaction parameters, 
based on the Boltzmann weight. 
Discrepancies from experimental data lead to empirically consider 
temperature dependence of the parameters, but these modifications 
keep unchanged the exponential nature of the equations.
We replace the Boltzmann weight by the nonextensive Tsallis weight,
and generalize three models for nonelectrolyte solutions that use 
the local composition hypothesis, namely Wilson's, NRTL, and UNIQUAC models.
The proposed generalizations present a nonexponential dependence 
on the temperature, 
and relies on a theoretical basis of nonextensive statistical mechanics.
The $q$-models present one extra binary parameter $q_{ij}$,
that recover the original cases in the limit $q_{ij} \to 1$.
Comparison with experimental data is illustrated with two examples of
the activity coefficient of ethanol, 
infinitely diluted in toluene, and in decane.
\end{abstract}


\maketitle

\section{Introduction}

The basic hypothesis of local composition models, 
empirically introduced by Wilson \cite{wilson}, 
and later used in other theories that followed, 
e.g. NRTL \cite{nrtl}, UNIQUAC \cite{uniquac},
assumes that the composition in the vicinity of a central molecule 
differs from the bulk composition, and this local inhomogeneity 
strongly affects thermodynamic properties of the solution.
The description of binary and multicomponent mixtures depends on 
interaction parameters, that are estimated from experimental data, 
and they usually have the general form
\begin{equation}
\label{eq:A1}
 A_{ij}=\exp\left( - \frac{\Delta a_{ij}}{RT}\right),
\end{equation}
with $\Delta a_{ij}=a_{ij}-a_{jj}$, and
$a_{ij}$ is a molar potential energy of interaction between species 
$i$ and $j$, with $a_{ji}=a_{ij}$. 
For a comprehensive approach to theory of solutions, and fluid-phase
equilibria in general, see \cite{prausnitz}.
%
$\Delta a_{ij}$ is originally assumed to be constant.
Extensions of the models relax this hypothesis, and consider 
$\Delta a_{ij}=\Delta a_{ij}(T)$,
according to various functions, for instance, a linear relation \cite{nrtl}, 
\begin{equation}
\label{eq:thomsen}
 \Delta a_{ij}=\Delta a_{ij,0}+\Delta a_{ij,1}(T-T_0),
\end{equation}
the inverse of the absolute temperature \cite{anderson-prausnitz},
\begin{equation}
\label{eq:anderson}
 \Delta a_{ij}=\Delta a_{ij,0}+\frac{\Delta a_{ij,1}}{T},
\end{equation}
a combination of both linear and inverse relation \cite{escobedo-sandler},
\begin{equation}
 \Delta a_{ij}=\Delta a_{ij,0}+\Delta a_{ij,1} T
 +\frac{\Delta a_{ij,2}}{T},
\end{equation}
or, else \cite{demirel-gecegormez-paksoy-p1:1992},
\begin{equation}
\label{eq:schmidt}
 \Delta a_{ij}=\Delta a_{ij,0}+\frac{\Delta a_{ij,1}}{T-T_0},
\end{equation}
or even with a logarithmic term \cite{larsen:1987}:
\begin{eqnarray}
\label{eq:larsen:1987}
 \Delta a_{ij}=\Delta a_{ij,0}+\Delta a_{ij,1}(T-T_0) 
 \nonumber \\
 +\Delta a_{ij,2} \left(T\ln \frac{T_0}{T} + T - T_0 \right).
\end{eqnarray}
Different thermodynamic properties
(heat of mixing, heat capacity, limiting activity coefficient,
vapor-liquid equilibrium, liquid-liquid equilibrium, etc.), and/or 
different chemical systems (presence of alcohols, hydrogen bonding, etc.),
may require different expressions for $\Delta a_{ij}(T)$. 
The use of these equations also varies according to the considered model.
Despite of the variety of functional forms, the exponential nature of the 
parameter, that stems from the Boltzmann's weight, is kept unchanged 
by all models.
%
Our proposal is to consider the nonextensive Tsallis weight,
as a replacement for the Boltzmann weight.
The departure from the exponential behavior is, thus, intrinsically
originated from the distribution, and not due to empirical modifications 
on the temperature dependence of the parameters.

The paper is divided as follows: 
in Section \ref{sec:next} we briefly present basic concepts of 
nonextensive statistical mechanics, that will be used later.
Section \ref{sec:local_composition} introduces the $q$-local composition model,
with the nonextensive weight.
Section \ref{sec:q-models} applies the nonextensive local composition 
to Wilson's, NRTL and UNIQUAC models.
Section \ref{sec:temperature_effects} illustrates the effect of the
nonextensive parameter $q_{ij}$ on the temperature, with instances
of activity coefficient at infinite dilution.  
Finally, 
Section \ref{sec:final} is dedicated to our conclusions and final remarks.

\section{\label{sec:next}Brief remarks on nonextensive statistical mechanics}

Despite the outstanding success of the Boltzmann-Gibbs statistical mechanics 
(BG), there are systems that are not properly described by the equations that
emerge from this formalism.
Along the last two decades there has been continuously and increasingly
developed the nonextensive statistical mechanics.
Its starting point is the generalization of the concept of entropy
(the Tsallis entropy) \cite{ct:1988},
\begin{equation}
\label{eq:sq}
 S_q=k\frac{1-\sum_i^W p_i^q}{q-1},
\end{equation}
with $p_i$ the probability of the microscopic state $i$, $W$ is the number
of microscopic states, $k$ is a positive constant, 
and $q$ is the entropic index.
If $q\to1$, Eq.~(\ref{eq:sq}) recovers the BG entropy,
$S_{BG}\equiv S_1=-k\sum_i^W p_i \ln p_i$, 
and thus $S_q$ is a generalization of $S_{BG}$.
%
Legendre transforms are preserved in nonextensive statistical mechanics 
\cite{curado-tsallis}.
%
The $q$-entropy in the microcanonical ensemble (maximization of $S_q$ with 
equiprobabilities) \cite{ct:1988} is
$S_q = k \ln_q W$, where the $q$-logarithm is defined as \cite{ct:quimicanova}
\begin{equation}
\label{eq:qlog}
 \ln_q x \equiv \frac{x^{1-q}-1}{1-q}.
\end{equation}
The celebrated equation, engraved in Boltzmann's tombstone, 
$S=k\ln W$, is recovered at $q\to1$.
Maximization of $S_q$, Eq.~(\ref{eq:sq}), with the constraint of constant 
generalized mean energy, leads to the canonical ensemble distribution 
for the energy
(see \cite{curado-tsallis,ct-mendes-plastino} for details),
\begin{equation}
\label{eq:tsallis-factor}
 p(x) \propto \exp_q(-\beta_q E_i),
\end{equation}
where $\beta_q$ is is the Lagrange parameter, that is related to 
the inverse temperature ($\beta_1=1/(kT)$ in the BG formalism), 
$E_i$ is the energy of the $i$-th state, and the $q$-exponential is precisely 
the inverse function of the $q$-logarithm, Eq.~(\ref{eq:qlog}),
\begin{equation}
\label{eq:qexp}
 \exp_q x = [1+(1-q)x]_+^{\frac{1}{1-q}}.
\end{equation}
The symbol $[A]_+$ stands for $[A]_+=A$ if $A>0$, and $[A]_+\equiv0$ if $A\le0$.
Equation (\ref{eq:tsallis-factor}) is the Tsallis weight, 
that is a generalization of the Boltzmann weight.
The main difference between Tsallis and Boltzmann weights
is that the former presents power law tails 
(long-lasting for $q>1$, and abruptly vanishing for $q<1)$,
while the later has exponential tails.
Equations (\ref{eq:qlog}) and (\ref{eq:qexp}) 
present many similar properties of the usual logarithm and exponential, 
e.g.  $\ln_{q}1=0$ and $\exp_{q}0=1$, $\forall q$, and
the derivative of the $q$-exponential is given by
\begin{equation}
\label{eq:deqdx}
 \frac{d (\exp_q x)}{dx} = (\exp_q x)^q.
\end{equation}
In general, $(\exp_q x)^a \ne \exp_q (ax)$, except for $q=1$.
The $q$-exponential and the $q$-logarithm functions lead to 
a nondistributive $q$-deformed algebra \cite{wang},\cite{epb:qalgebra}. 
Generalized algebraic operations 
($q$-addition $x \oplus_q y$, $q$-difference $x \ominus_q y$,
$q$-product $x \otimes_q y$, $q$-ratio $x \oslash_q y$) are defined as
\begin{equation}
\label{eq:qsum}
 x\oplus_q y \equiv x+y+(1-q)xy,
\end{equation}
\begin{equation}
\label{eq:qdiff}
 x\ominus_q y \equiv \frac{x-y}{1+(1-q)y}, \quad (y\ne \frac{1}{q-1}),
\end{equation}
\begin{equation}
\label{eq:qproduct}
 x\otimes_q y \equiv \left[x^{1-q}+y^{1-q}-1\right]^{\frac{1}{1-q}}_+
 \quad (x,y>0),
\end{equation}
\begin{equation}
\label{eq:qratio}
 x\oslash_q y \equiv \left[x^{1-q}-y^{1-q}+1\right]^{\frac{1}{1-q}}_+
 \quad (x,y>0).
\end{equation}
With these $q$-operations, the $q$-exponential follows the properties:
\begin{eqnarray}
  \begin{array}{lll}
  \exp_q x \exp_q y          &=&\exp_q(x\oplus_qy), 
  \\
  \exp_q x / \exp_q y         &=&\exp_q(x\ominus_qy),
  \\
  \exp_q x \otimes_q \exp_q y &=&\exp_q(x+y),
  \\
  \exp_q x \oslash_q \exp_q y &=&\exp_q(x-y).
  \end{array}
\end{eqnarray}
The $q$-algebra has been applied in different contexts within nonextensive
statistical mechanics. 
The $q$-product has been used in the generalizations of 
the central limit theorem and the Fourier transform 
\cite{ct:milan},\cite{umarov-tsallis-steinberg},\cite{jauregui-tsallis:2011},
and there are evidences that it is connected to $q$-Gaussian distributions
$p(x)=A(q)\sqrt{\beta}\exp_q(-\beta x^2)$ \cite{moyano-tsallis-gellmann}.
Some properties of $q$-functions and $q$-algebra may be found at 
\cite{yamano:2002},%
\cite{naudts:2002},%
\cite{epb:jmp2008},%
\cite{ct:springer},
and references therein.
%

Nonextensive statistical mechanics is expected to be valid 
in a variety of situations:
systems with long range interactions, long term memory, fractal structure,
break of ergodicity, quasi-stationary states,
or other features that characterize complex behavior.
Let us briefly elaborate on the range of interactions,
and on the nature of the quasi-stationary states,
following the lines of \cite{ct:bjp}.
%
We can generally consider that interactions decay with distance $r$
as $1/r^\alpha$. If $\alpha<3$, the interaction is long-ranged;
$\alpha=1$, for Coulomb and gravitation interactions, are typical examples.
Such systems may exhibit negative specific heat, e.g. \cite{epb:lj}.
If $\alpha>3$, the interaction is short-ranged,
e.g. van der Waals $\alpha=6$ interactions.
Dipole interactions are at the threshold ($\alpha=3$).
See \cite{ct:bjp} for weak and strong violation of BG, and more details.

Thermodynamic equilibrium is concerned about two limits: 
the time limit (time $t\to \infty$, related to {\em equilibrium}) 
and the macroscopic limit (number of particles $N\to\infty$, 
related to {\em thermodynamics}).
For simple systems, the order in which these limits are taken is irrelevant,
but for complex systems, these limits may not commute.
For certain classes of complex systems,
if the time limit is taken first, and then the macroscopic limit
($\lim_{N\to\infty}\lim_{t\to\infty}f(t,N)$ of a dynamical function $f(t,N)$),
the system is characterized by Boltzmann equilibrium distributions, thus $q=1$.
But if the limits are taken in the reverse order 
($\lim_{t\to\infty}\lim_{N\to\infty}f(t,N)$),
the system may achieve a quasi-stationary metaequilibrium state, 
according to its initial conditions,
that is possibly described by nonextensive distributions. 
This hypothesis was conjectured in 1999 by Tsallis \cite{ct:bjp}, 
and it was computationally verified for conservative long-range interacting 
systems, e.g.
\cite{Latora-Rapisarda-Tsallis:2001},%
\cite{Latora-Rapisarda-Tsallis:2002},%
\cite{Tsallis-Rapisarda-Pluchino-Borges:2007}.

We address  some examples that follow nonextensive behavior.
The rate of re-association of CO with Myoglobin dissolved in glycerol-water 
solutions, after being photo-dissociated,
was found to be described by functions that are connected to 
nonextensive statistical mechanics \cite{ct-bemski-mendes}.
This was supposed to be related to the path CO molecules have to perform 
in a fractal-like structure though the interior of folded proteins. 
$q$-Gaussian distributions were applied to polymeric networks, 
when finite chain effects are relevant \cite{malacarne-polymers}.
Nonextensive distributions of velocity of monomers during the relaxation 
process were found by molecular dynamics simulations of polymer chains 
and Lennard-Jones molecules \cite{hu-ma}.
$q$-Exponential functions were used to describe diatomic potential energy 
curves, particularly for H$_2^+$ and Li$_2$, and vibrational spectra and 
spectroscopic constants were found to be in good agreement with experimental 
data \cite{heibbe}.
Arrhenius law has recently been generalized by the use of the $q$-exponential, 
and agreement with experiment was found in plant respiration rates, 
bacterial gliding, and tunneling in the F + H$_2$ reaction \cite{qArrhenius}.
Theoretical and experimental aspects, and the historical development 
of nonextensive statistical thermodynamics, and many examples,
may be found in \cite{ct:springer}.
See also \cite{naudts:gt} for a theoretical treatment 
of generalized thermostatistics.

\section{\label{sec:local_composition}Nonextensive local composition}

The basic assumption of the local composition theory,
as introduced by Wilson \cite{wilson}, is that, 
due to differences in molecular sizes and in intermolecular interactions, 
the ratio of  the number of molecules of species $i$ and $j$, in the vicinity 
of a central molecule $j$, differs from that of the whole solution, according to
$x_{ij}/x_{jj}=(x_i/x_j) 
               (\exp(-\frac{a_{ij}}{RT}) / \exp(-\frac{a_{jj}}{RT})),
$
where $x_i$ is the bulk mole fraction of species $i$, 
and $x_{ij}$ is the local mole fraction of species $i$ in the neighborhood 
of a molecule of species $j$.
Our assumption is to replace the Boltzmann weight 
by the nonextensive Tsallis weight. 
Besides, the $q$-product of probabilities yields nonextensive distributions 
(distributions remarkably close to $q$-Gaussians, to be more precise; 
see \cite{moyano-tsallis-gellmann},\cite{hilhorst:2007},\cite{ct:springer}).
The $q$-product, and the $q$-ratio, of Tsallis weights has recently been
shown to be related to the locality of a generalized master equation
\cite{martinez-metropolis}.
This inspires us to use the $q$-ratio, in the generalization of 
the local composition hypothesis: for a central molecule $j$, 
\begin{equation}
\frac{x_{ij}}{x_{jj}}=
\label{eq:xijxjj}
                       \frac{x_i}{x_j}
                       \exp_{q_{ij}}\left(-\frac{a_{ij}}{RT}\right) 
                       \oslash_{q_{ij}}
                       \exp_{q_{ij}}\left(-\frac{a_{jj}}{RT}\right).
\end{equation}
Substitution of Eq.~(\ref{eq:xijxjj}) in the normalization condition 
$\sum_i^c x_{ij}=1$ ($c$ is the number of chemical species) results
\begin{equation}
\label{eq:xij}
 x_{ij}=\frac{x_i A_{q,ij}}{\sum_k^c x_k A_{q,kj}},
\end{equation}
with the parameter $A_{q,ij}$ given by%
\footnote{We adopt the symbol $A_{q,ij}$, instead of
$A_{q_{ij},ij}$, to avoid unnecessary heavy notation.}
\begin{equation}
\label{eq:Aq}
 A_{q,ij} \equiv \exp_{q_{ij}}\left(-\frac{\Delta a_{ij}}{RT}\right),
\end{equation}
with $\Delta a_{ij}=a_{ij}-a_{jj}$. 
Symmetry of the interactions implies $a_{ji}=a_{ij}$,
and we assume, for simplicity, $q_{ji}=q_{ij}$.
The limiting case $q_{ij}\to 1$ recovers the usual parameter 
$A_{ij}\equiv A_{1,ij}$, Eq.~(\ref{eq:A1}).
Note that, according to Eq.~(\ref{eq:deqdx}),
\begin{equation}
\label{eq:dAqdT}
 \frac{dA_{q,ij}}{dT}= A_{q,ij}^{q_{ij}} \frac{\Delta a_{ij}}{RT^2}.
\end{equation}

\begin{figure}[htb]
\begin{center}
 \includegraphics[width=0.75\columnwidth,keepaspectratio,clip]{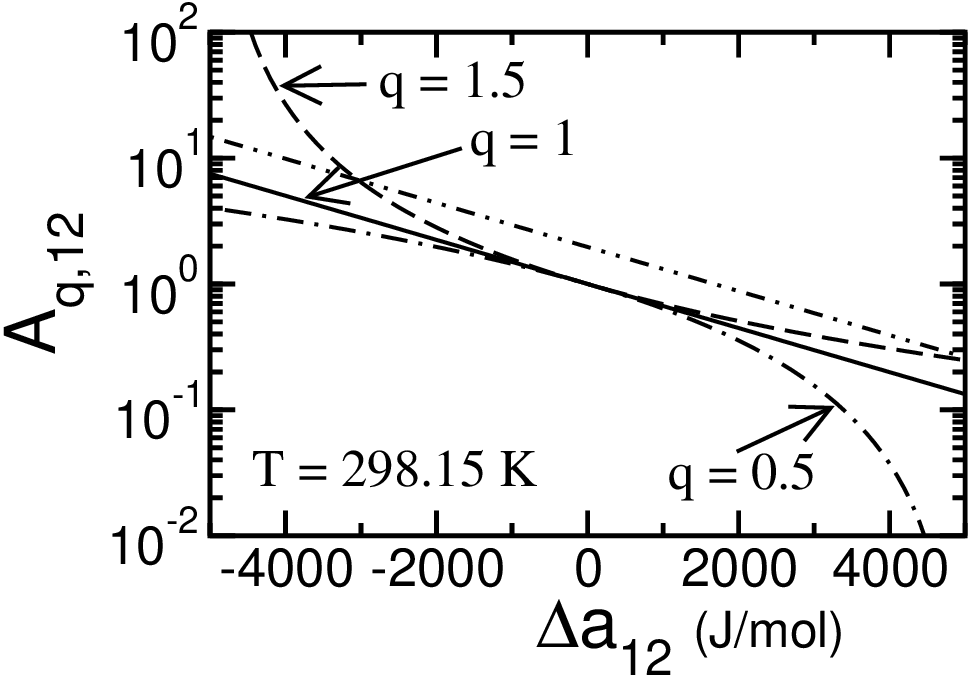}
\end{center}
\caption{\label{fig:A12-a12}%
         Parameter $A_{q,12}$, Eq.~(\protect\ref{eq:Aq}), as a function of 
         the binary interaction parameter $\Delta a_{12}$ (with $T=298.15$~K).
         $q_{12}=0.5$ (dot-dashed), $q_{12}=1$ (solid), 
         and $q_{12}=1.5$ (dashed).
         Dot-dot-dashed curve uses Eq.~(\ref{eq:anderson}) with 
         $\Delta a_{12,1}=5\times10^5$~J~K/mol.
}
\end{figure}

\begin{figure}[htb]
\begin{center}
 \includegraphics[width=0.75\columnwidth,keepaspectratio,clip]{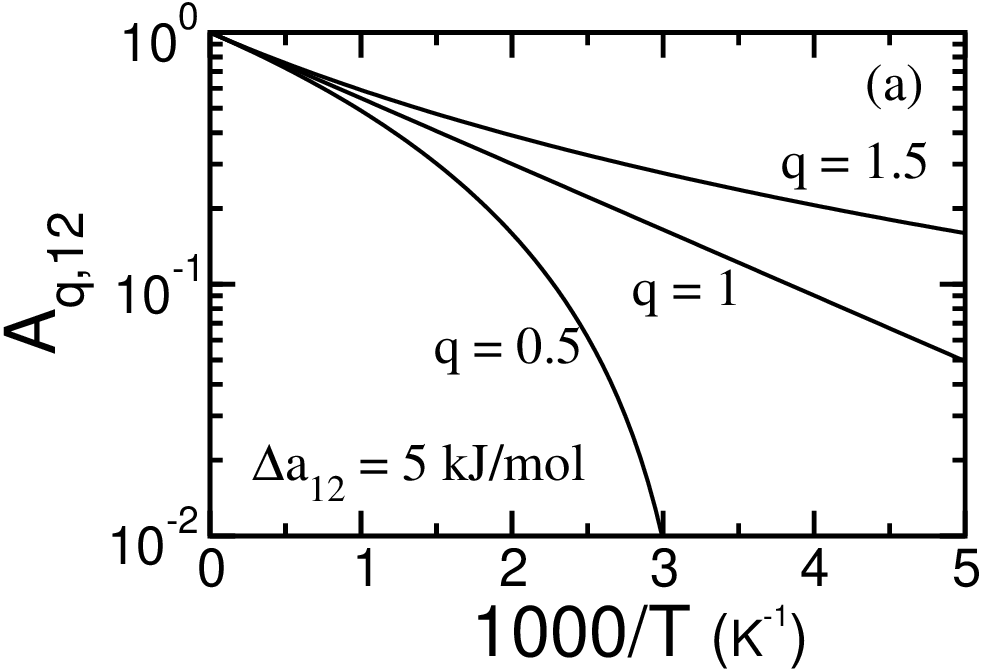}
 \includegraphics[width=0.75\columnwidth,keepaspectratio,clip]{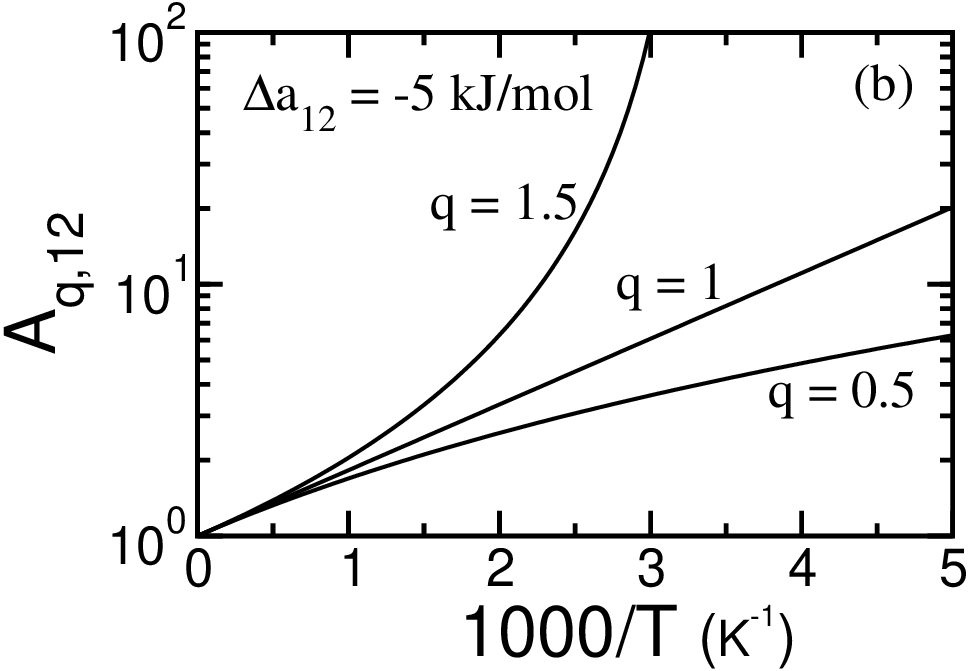}
\end{center}
\caption{\label{fig:A12-1suT}%
         Parameter $A_{q,12}$, Eq.~(\protect\ref{eq:Aq}),
         as a function of the inverse temperature (conveniently scaled),
         for two typical values of $\Delta a_{12}$.
         The ordinary $q_{12}=1$ model appears as straight lines 
         in these semi-log plots. 
         $q_{12}>1$ ($q_{12}<1$) presents positive (negative) concavity.
}
\end{figure}

Figures \ref{fig:A12-a12} and \ref{fig:A12-1suT} show the behavior of the 
parameter $A_{q,12}$, Eq.~(\ref{eq:Aq}), as a function of $\Delta a_{12}$, 
and the scaled inverse temperature, respectively. 
The usual case $q_{12}=1$ appears as straight lines in these semi-logarithmic 
plots, and $q_{12}>1$ ($q_{12}<1$) presents positive (negative) concavity.
Equation (\ref{eq:anderson}) is also displayed in Fig.~\ref{fig:A12-a12}, 
for comparison.
The effect of the parameter $\Delta a_{12,1}$ of Eq.~(\ref{eq:anderson})
is simply to shift the $q_{12}=1$ curve, 
but it remains a straight line, once it relies on the Boltzmann weight.

\section{\label{sec:q-models}$q$-Wilson's, $q$-NRTL, and $q$-UNIQUAC models}

\subsection{$q$-Wilson's model}

The excess molar Gibbs free energy for the Wilson's model \cite{wilson}
is an empirical modification of the Flory-Huggins' equation
(see \cite{prausnitz} for details and additional references):
\begin{equation}
\label{eq:ge-qwilson}
 \frac{g^E}{RT} = \sum_i^c x_i \ln \frac{\xi_{ii}}{x_i},
\end{equation}
where $\xi_{ii}$ is the local volumetric fraction of the component $i$ 
in the neighborhood of another molecule $i$ (Flory-Huggins' model for athermal 
polymeric solutions uses the global segment fraction $\xi_i$ in the place of 
$\xi_{ii}$, in Eq.~(\ref{eq:ge-qwilson})).
The local volumetric fraction is given by
\begin{equation}
\label{eq:xi_ii}
 \xi_{ii}=\frac{x_{ii} v_i}{\sum_j^c x_{ji} v_j},
\end{equation}
where the molar volume of liquid $i$, $v_i$, is taken as a measure of its 
molecular volume.

Substitution of Eq.s~(\ref{eq:xij}) and (\ref{eq:xi_ii}) 
in Eq.~(\ref{eq:ge-qwilson}), leads to the nonextensive generalization 
of the Wilson's model, and the excess molar Gibbs free energy is
\begin{equation}
 \frac{g^E}{RT}=-\sum_i^c x_i \ln\left(\sum_j^c x_j \Lambda_{q,ij}\right),
\end{equation}
with $\Lambda_{q,ij}=(v_j/v_i)A_{q,ij}$.
The activity coefficient is formally given by the same expression
of the original ($q_{ij}=1$) model, but with the $q$-parameter $\Lambda_{q,ij}$:
\begin{equation}
 \ln \gamma_i = -\ln\left(\sum_j^c x_j \Lambda_{q,ij}\right) + 1 
         - \sum_j^c \frac{x_j \Lambda_{q,ji}}{\sum_k^c x_k \Lambda_{q,jk}}.
\end{equation}
The excess molar enthalpy (heat of mixing)
$h^E=\frac{\partial (g^E/T)}{\partial (1/T)}$ 
is given by (see Eq.~(\ref{eq:dAqdT}))
\begin{equation}
h^E = \sum_i^c \frac{x_i}{\sum_j^c x_j \Lambda_{q,ij}}
\sum_k^c \left(\frac{v_k}{v_i}\right)^{\!\!1-q_{ik}} x_k \Lambda^{q_{ik}}_{q,ik} 
\Delta a_{ik}.
\end{equation}

\subsection{$q$-NRTL model}

The nonextensive generalization of the NRTL model follows the same
lines of \cite{nrtl}, with Eq.~(\ref{eq:xijxjj}).
The excess molar Gibbs free energy is given by
\begin{eqnarray}
\label{eq:ge-nrtl} 
 g^E &=& \sum_i^c x_i \sum_j^c x_{ji} \Delta a_{ji} \nonumber \\
     &=& \sum_i^c x_i \frac{\sum_j^c x_j A_{q,ji} \Delta a_{ji}}
                           {\sum_k^c x_k A_{q,ki}}.
\end{eqnarray}
The NRTL model introduces a nonrandomness parameter $\alpha_{ij}=\alpha_{ji}$,
so the parameter $A_{q,ij}$ is given by a variation of Eq.~(\ref{eq:Aq}), 
namely
$ A_{q,ij} = \exp_{q_{ij}}(-\alpha_{ij}\frac{\Delta a_{ij}}{RT})$.
The expression for the activity coefficient of a component $i$ is formally
the same as the original ($q_{ij}=1$) NRTL model, just replacing 
the usual parameter $A_{ij}$ by $A_{q,ij}$.
This procedure is not valid to find the expressions of excess molar entropy 
and excess molar enthalpy, due to Eq.~(\ref{eq:dAqdT}).

\subsection{$q$-UNIQUAC model}

The UNIQUAC model \cite{uniquac} is based on the local composition hypothesis, 
but it replaces the local and global mole fractions in Eq.s~(\ref{eq:xijxjj})
and (\ref{eq:xij}) 
by the local and global surface fractions $\theta_{ij}$ and $\theta_i$
(the global surface fraction is defined as
$\theta_i = x_i \tilde{q}_i / (\sum_j^c x_j \tilde{q}_j)$, 
where $\tilde{q}_i$ is the surface parameter%
\footnote{%
We use the notation $\tilde{q}_i$ to avoid confusion with 
the nonextensive parameter $q_{ij}$.
Sometimes there appear two surface parameters, 
$\tilde{q}$ and $\tilde{q}'$, one of them is used in the combinatorial term 
and the other in the residual term of the UNIQUAC model 
\protect\cite{anderson-prausnitz}. 
We consider $\tilde{q}=\tilde{q}'$, for simplicity.%
}).
The interaction parameter $a_{ij}$ is considered as a measure of the 
internal energy ($a_{ij}=u_{ij}$), and the excess molar internal energy 
is then given by
\begin{equation}
\label{eq:ue}
 u^E = \sum_i^c x_i \tilde{q}_i \sum_j^c \theta_{ji} \Delta u_{ji}.
\end{equation}
The excess molar Gibbs free energy is found by the approximate relation
\begin{equation}
\label{eq:ge-ue}
 g^E/T \approx \int_0^{1/T} u^E d(1/T),
\end{equation}
that, for the multicomponent case, shall be numerically integrated.
The residual contribution for the activity coefficient of component $i$, 
for a multicomponent mixture, may be found by the numerical integration of 
\begin{equation}
\label{eq:lngamma}
 \ln\gamma_i^{res} = \int_0^{1/T}\bar{u}_i^E d(1/T),
\end{equation}
with the partial molar excess internal energy given by
\begin{eqnarray}
\label{eq:ubari}
 \bar{u}_i^E = \tilde{q}_i \sum_j^c \frac{\theta_j A_{q,ji}\Delta u_{ji}}
                                         {\sum_k^c \theta_k A_{q,ki}}
             + \tilde{q}_i \sum_j^c \frac{\theta_j A_{q,ij}\Delta u_{ij}}
                                         {\sum_k^c \theta_k A_{q,kj}}
             \nonumber \\
             - \tilde{q}_i \sum_j^c \frac{\theta_j A_{q,ij}}
                                         {(\sum_k^c \theta_k A_{q,kj})^2}
                        \textstyle\sum_k^c \theta_k A_{q,kj}\Delta u_{kj}.
\end{eqnarray}

Analytical solution for the binary case is as follows:
substitution of Eq.~(\ref{eq:ue}) in Eq.~(\ref{eq:ge-ue}), 
and Eq.~(\ref{eq:ubari}) in Eq.~(\ref{eq:lngamma}),
with $c=2$, lead to (see Eq.~3.194~5 and Eq.~3.194~1 of \cite{gradshteyn})
\begin{eqnarray}
 \int_0^{x} 
 \frac{c\exp_q(-cx)}{a+b\exp_q(-cx)} dx =
      -\frac{1}{a} 
      \int_1^\tau \frac{\tau'^{1-q}}{1+\frac{b}{a}\tau'} d\tau'
      \nonumber \\
      = 
      -\frac{1}{a(2-q)} \left[\tau^{2-q} \chi_q\left(\frac{b}{a}\tau\right)
      - \chi_q\left(\frac{b}{a}\right)\right],
\end{eqnarray}
\begin{eqnarray}
 \int_0^{x} 
 \frac{c\exp_q(-cx)}{[a+b\exp_q(-cx)]^2} dx =
      -\frac{1}{a^2} 
      \int_1^\tau \frac{\tau'^{1-q}}{(1+\frac{b}{a}\tau')^2} d\tau'
      \nonumber \\
      = 
      -\frac{1}{a^2(2-q)} \left[\tau^{2-q} \psi_q\left(\frac{b}{a}\tau\right)
      - \psi_q\left(\frac{b}{a}\right)\right],
\end{eqnarray}
where we have used the change of variables $\tau=\exp_q(-cx)$, and 
$\int_1^\tau f(\tau')d\tau'=\int_0^\tau f(\tau')d\tau'-\int_0^1 f(\tau')d\tau'$,
with 
\begin{equation}
 \chi_q(x)=\,_2F_1(1,2-q;3-q;-x), 
\end{equation}
\begin{equation}
 \psi_q(x)=\,_2F_1(2,2-q;3-q;-x), 
\end{equation}
$q<2$, 
and $_2F_1(\alpha,\beta;\gamma;x)$ is the hypergeometric function,
resulting the following analytical expressions for the residual contributions,
$g^E_{res}$ and $\ln\gamma_i^{res}$:
\begin{widetext}
\begin{eqnarray}
\label{eq:ge-bin}
 \left(\frac{g^E}{RT}\right)_{res} = 
 - \frac{x_1 \tilde{q}_1}{(2-q_{_{12}})} \frac{\theta_2}{\theta_1}
 \left[
   \tau_{q,21}^{2-q_{_{12}}}
   \chi_{q_{_{12}}}\left(\frac{\theta_2}{\theta_1}\tau_{q,21}\right)
   -
   \chi_{q_{_{12}}}\left(\frac{\theta_2}{\theta_1}\right)
 \right]
\nonumber \\
 - \frac{x_2 \tilde{q}_2}{(2-q_{_{12}})} \frac{\theta_1}{\theta_2}
 \left[
   \tau_{q,12}^{2-q_{_{12}}}
   \chi_{q_{12}}\left(\frac{\theta_1}{\theta_2}\tau_{q,12}\right)
   -
   \chi_{q_{12}}\left(\frac{\theta_1}{\theta_2}\right)
 \right],
\end{eqnarray}
\begin{eqnarray}
\label{eq:lngamma-bin}
 \ln \gamma_i^{res} =
 - \frac{\tilde{q}_i}{(2-q_{_{12}})} \frac{\theta_j}{\theta_i}
 \left[
   \tau_{q,ji}^{2-q_{_{12}}}
   \chi_{q_{_{12}}}\left(\frac{\theta_j}{\theta_i}\tau_{q,ji}\right)
   -
   \chi_{q_{_{12}}}\left(\frac{\theta_j}{\theta_i}\right)
 \right]
\nonumber \\
 + \frac{\tilde{q}_i}{(2-q_{_{12}})} \frac{\theta_j}{\theta_i}
 \left[
   \tau_{q,ji}^{2-q_{_{12}}}
   \psi_{q_{_{12}}}\left(\frac{\theta_j}{\theta_i}\tau_{q,ji}\right)
   -
   \psi_{q_{_{12}}}\left(\frac{\theta_j}{\theta_i}\right)
 \right]
\nonumber \\
 - \frac{\tilde{q}_i}{(2-q_{_{12}})} 
 \left[
   \tau_{q,ij}^{2-q_{_{12}}}
   \psi_{q_{_{12}}}\left(\frac{\theta_i}{\theta_j}\tau_{q,ij}\right)
   -
   \psi_{q_{_{12}}}\left(\frac{\theta_i}{\theta_j}\right)
 \right],
\end{eqnarray}
\end{widetext}
with $q_{12}<2$, 
$\tau_{q,ij}\equiv A_{q,ij}=\exp_{q_{ij}}(-\Delta u_{ij}/(RT))$,
and ($i=1$, $j=2$) or ($i=2$, $j=1$).
The limiting case ${q_{12}}\to1$, with 
\mbox{$\chi_1(x)=x^{-1}\ln(1+x)$}, and 
\mbox{$\psi_1(x)=(1+x)^{-1}$}
(see Eq.~9.121~6 and Eq.~9.121~16 of \cite{gradshteyn},
and Eq.~15.3.15 of \cite{abramowitz}), recover the usual expressions.

It is possible to find an expression for the residual contribution of the
activity coefficient of species $i$, in a binary solution, different from, 
but equivalent to, Eq.~(\ref{eq:lngamma-bin}), with the partial molar excess 
Gibbs free energy taken from Eq.~(\ref{eq:ge-bin}), and
$\ln \gamma_i^{res}=(\bar{g}_i^E)_{res}/(RT)$, where it is necessary to use
the derivative of $\chi_q(x)$ (see Eq.~15.2.1 of \cite{abramowitz}), 
$d\chi_q(x)/dx = -(2-q)/(3-q) \,_2F_1(2,3-q;4-q;-x).$

The combinatorial contribution of the UNIQUAC model 
for the multicomponent case,
that is the Guggenheim expression for athermal mixtures, 
remains unchanged in the nonextensive generalization,
consistent with the lower limit of Eq.~(\ref{eq:ge-ue})
\cite{uniquac}:
\begin{eqnarray}
 \left(\frac{g^E}{RT}\right)_{comb} = 
   \sum_i^c x_i \ln\frac{\phi_i}{x_i}
    + \frac{z}{2} \sum_i^c \tilde{q}_i x_i \ln \frac{\theta_i}{\phi_i},
\end{eqnarray}
\begin{eqnarray}
 \ln \gamma_i^{comb} = \ln\frac{\phi_i}{x_i}
    + \frac{z}{2} \tilde{q}_i \ln \frac{\theta_i}{\phi_i}
    + l_i - \frac{\phi_i}{x_i} \sum_j^c x_j l_j
\end{eqnarray}
where the volume fraction $\phi_i = x_i r_i / (\sum_j^c x_j r_j)$, 
$r_i$ is the volumetric parameter of molecule $i$, 
$l_i=(z/2)(r_i-\tilde{q}_i)-(r_i-1)$,
and $z$ is the coordination number, usually assumed $z=10$.
The complete expression for the excess molar Gibbs free energy comprises 
the two contributions, $g^E = g^E_{res} + g^E_{comb}$.

\section{\label{sec:temperature_effects}%
Activity coefficient at infinite dilution}

\begin{figure}[htb]
\begin{center}
 \includegraphics[width=0.75\columnwidth,keepaspectratio,clip]{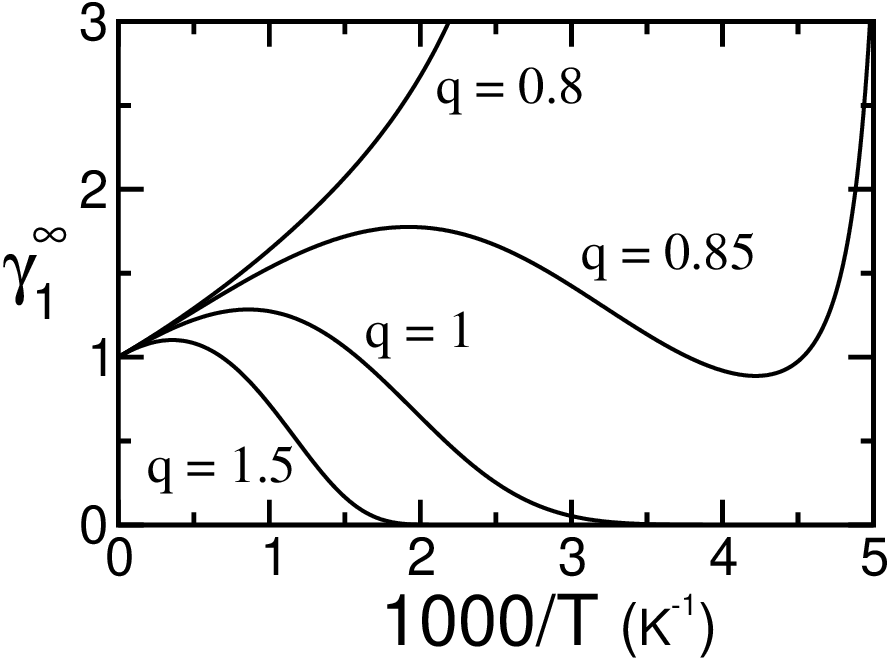}
\end{center}
\caption{\label{fig:ginfty}%
         $\gamma_1^\infty$ according to $q$-Wilson's model,
         as a function of the inverse temperature,
         for typical values of the binary parameters: 
         $\Delta a_{12}=10^4$~J/mol and $\Delta a_{21}=-5600$~J/mol
         ($v_1=v_2$, values of $q_{12}$ are indicated). 
         Curve for $q_{12}=0.85$ displays a maximum and a minimum.
         The positions of the extreme points and the values of $\gamma_1^\infty$
         may be changed by adjusting the parameters to other values; 
         the particular values of the parameters were chosen 
         to make the figure visually good.
}
\end{figure}

We have chosen the activity coefficient at infinite dilution, that is a 
property that depends only on the temperature, to illustrate our proposal. 
We have also chosen the $q$-Wilson's model to illustrate the effect of the 
nonextensive parameter $q_{ij}$ on the temperature dependence,
because it is the simplest model, if compared to $q$-NRTL and $q$-UNIQUAC.
The activity coefficient at infinite dilution for the $q$-Wilson's model 
is 
$\ln \gamma_1^\infty = \ln \Lambda_{q,12} - \Lambda_{q,21} + 1.$
Fig.~\ref{fig:ginfty} shows curves for different values of $q_{12}$
(all curves use the same parameters $\Delta a_{12}$ and $\Delta a_{21}$).
$q_{12}=0.85$ exhibits a maximum and a minimum in $\gamma_1^\infty$,
while $q_{12}=1$ can only display a maximum.

\begin{figure}[htb]
\begin{center}
 \includegraphics[width=0.75\columnwidth,keepaspectratio,clip]{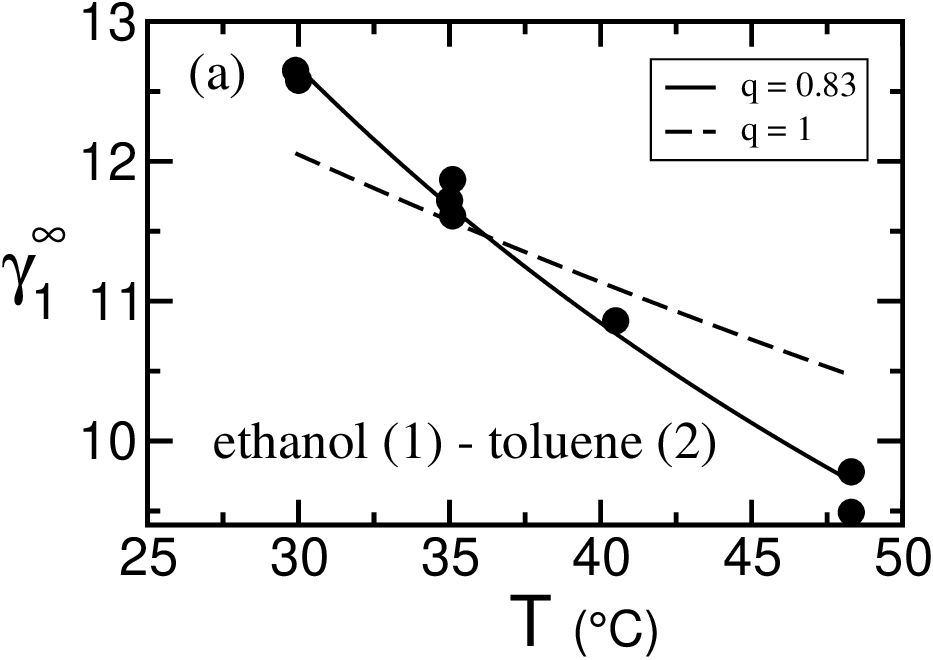}
 \includegraphics[width=0.75\columnwidth,keepaspectratio,clip]{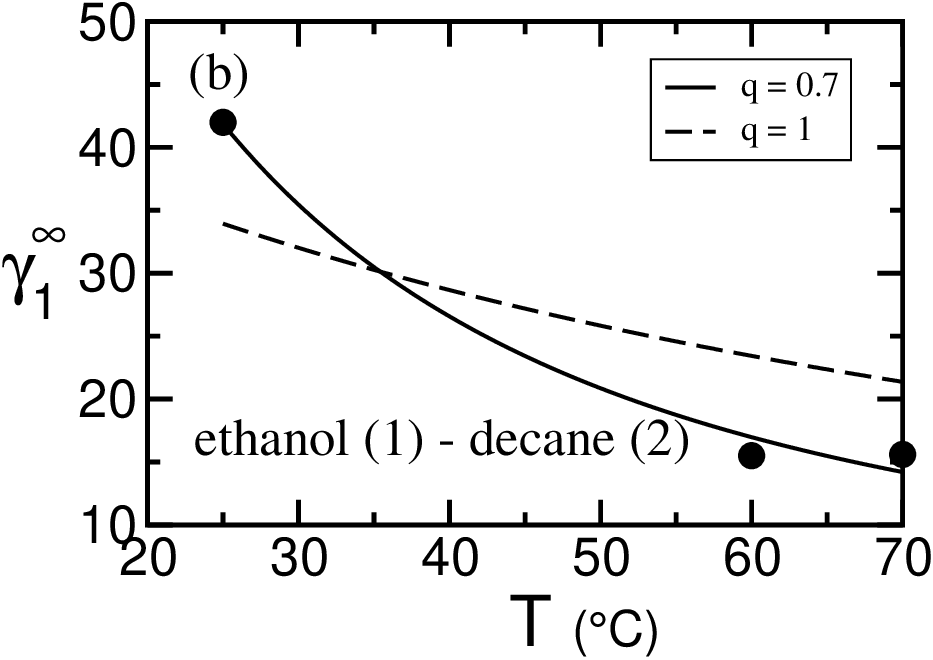}
\end{center}
\caption{\label{fig:ethanol}%
         Activity coefficient at infinite dilution of ethanol 
         in the solvents toluene (a), and decane (b).
         Experimental data from \protect\cite{dechema}. 
         Dashed lines are the best fittings for the original 
         ($q_{12}=1$) Wilson's model, and
         solid lines are the best fittings for the $q$-Wilson's model.
         (a) Dashed line: 
             $\Delta a_{12}= 6880.0$~J/mol, $\Delta a_{21}=-582.0$~J/mol;
             solid line: 
         $q_{12}=0.83$, 
         $\Delta a_{12}=10267.0$~J/mol, $\Delta a_{21}=-5179.7$~J/mol.
         (b) Dashed line: 
             $\Delta a_{12}= 8773.6$~J/mol, $\Delta a_{21}=-26.6$~J/mol;
             solid line: 
         $q_{12}=0.7$, $\Delta a_{12}=6500.0$~J/mol, 
         $\Delta a_{21}=-2353.7$~J/mol.}
\end{figure}

We have fitted the $q$-Wilson's model to two examples of experimental
activity coefficient at infinite dilution:
ethanol infinitely diluted in toluene (Fig.~\ref{fig:ethanol}a),
and in decane (Fig.~\ref{fig:ethanol}b).
The examples are not properly described by the ordinary Wilson's model,
and the generalized $q$-Wilson's model is able to describe the data.
The data were taken from \cite{dechema} (Vol IX, Parts 3, 4)%
\footnote{%
Four references are reported in \protect\cite{dechema} for the system 
ethanol-toluene, and their fluctuations are significative.
To avoid such fluctuations, we have considered one single set of experimental 
data, measured with the dilutor technique: Ref.~10 of \cite{dechema},
Vol IX, Part 3, pp.~1292--1293 
\protect\cite{toluene-ethanol-ref10-dechema}.}.
The two examples present positive deviations from Raoult's law
($\gamma_1^\infty >1$) and decreasing $\gamma_1^\infty$ with the temperature.

\section{\label{sec:final}Final remarks}

The main goal of this paper is to introduce nonextensivity
in local composition models, used in theories of solutions. 
Current theories and models are based on Boltzmann's distribution 
and Boltzmann's weight. 
Deviations from Boltzmann's weight have been proposed, mainly on an 
empirical basis (see Eq.s~(\ref{eq:thomsen})-(\ref{eq:larsen:1987})).
The $q$-local composition hypothesis introduces one additional 
binary parameter that comes from the nonextensive theory,
and generalizes the temperature dependence of the models. 

As one increases the degree of freedom of a model by adding extra parameters, 
it is natural to expect a better optimization from a fitting procedure.
Sometimes new parameters are simply introduced to take advantage 
of the additional degrees of freedom, and then turn the fittings easier.
Nonextensive distributions do present a new parameter $q_{ij}$, and of course
the fittings are benefited from it.
But the extra parameter $q_{ij}$ lies on a theoretical background, 
and there are plenty of examples showing that the entropic index $q$
has a physical interpretation, expressing the degree of nonextensivity 
of the system \cite{ct:springer}.

We have shown two fittings for the activity coefficient at infinite dilution
for the $q$-Wilson's model. 
Expressions for other models ($q$-NRTL, $q$-UNIQUAC) were presented, 
but applications to large amounts of experimental data still remain 
to be explored.

We hope this work invites experimentalists to apply nonextensive local 
composition models to large experimental databases, 
and test their validity in the description, or predicting capability, 
of thermodynamic properties, in different temperatures.

\begin{acknowledgments} 
We thank Alexandre Souto Martinez for interesting remarks.
This work was partially supported by FAPESB, through the program PRONEX
(Brazilian funding agency).
\end{acknowledgments} 

\bibliography{qge}

\end{document}